\documentclass{aa}
\usepackage{graphicx,psfig}

\newcommand{\Msun} {M$_\odot$}
\newcommand{\Lsun} {L$_\odot$}

\newcommand{\Tstar} {T$_{\rm{eff}}$}
\newcommand{\Lstar} {L$_\star$}
\newcommand{\Mstar} {M$_\star$}

\newcommand{\um} {$\mu$m}

\newcommand{\Md} {M$_{d}$}
\newcommand{\Ri} {R$_{in}$}
\newcommand{\Rd} {R$_{d}$}
\newcommand{\amm} {$\alpha_{\rm{mm}}$}

\newcommand{\grad} {$^{\rm o}$}

\newcommand{\Mk} {M$_d^\kappa$}
\newcommand{\Mku} {M$_d \times(\kappa_{\rm{1mm}}/0.01)$}
\newcommand{\bmm} {$\beta$}
\newcommand{\bsd} {$\beta_{SD}$}
\newcommand {\amax} {a$_{max}$}
\newcommand {\amin} {a$_{min}$}
\newcommand{\Mratio} {M$_d$/M$_\star$}

\newcommand{\simless}{\mathbin{\lower 3pt\hbox
      {$\rlap{\raise 5pt\hbox{$\char'074$}}\mathchar"7218$}}} 
\newcommand{\simgreat}{\mathbin{\lower 3pt\hbox
     {$\rlap{\raise 5pt\hbox{$\char'076$}}\mathchar"7218$}}} 

\begin{document}

\title{A search for evolved dust in  Herbig Ae stars}

\author{
A. Natta\inst{1},
L. Testi\inst{1},
R. Neri \inst{2},
D. S.  Shepherd \inst{3},
D.J. Wilner \inst{4}
}
\institute{
    Osservatorio Astrofisico di Arcetri, INAF, Largo E.Fermi 5,
    I-50125 Firenze, Italy
\and
IRAM, 300 rue de la Piscine, F-38406 St Martin d'Heres, France
\and
National Radio Astronomy Observatory,P.O. Box O, Socorro, NM 87801
\and
Harvard-Smithsonian Center for Astrophysics, 60 Garden Street,
       Cambridge, MA 02138
}

\offprints{lt@arcetri.astro.it}
\date{Received ...; accepted ...}

\authorrunning{Natta et al.}
\titlerunning{Evolved dust in Herbig Ae stars}

\abstract{
%
We present observations of six isolated, pre-main-sequence, intermediate mass 
stars selected for shallow spectra at submillimeter wavelengths at 
1.3, 2.6, 7.0, and 36 millimeters from the IRAM PdBI and the VLA. 
We analyze the new observations of these stars (HD34282, HD35187, HD142666,
HD143006, HD150193, HD163296) together with similar observations of three 
additional stars from the literature (CQ Tau, UX Ori, TW Hya), in the context 
of self-consistent irradiated disk models. Our aim is
to constrain the wavelength 
dependence of the dust opacity and the total dust mass in the disks.
The shallow wavelength dependence of the opacity is confirmed and for a few
stars extended to significantly longer wavelengths.
For any plausible dust properties, this requires grain growth from 
interstellar sizes to maximum sizes of at least a few millimeters, and 
very likely to several centimeters or more. 
For four of the stars (HD34282, HD163296, CQ Tau, TW Hya), the millimeter 
emission has been spatially resolved, and the large disk radii ($>100$ AU) 
rule out that  high optical
depths play a role.
The mass of dust that has been processed into large grains is substantial, and 
in some cases implies a disk mass comparable to the mass of the central star.
}
\maketitle

\section{Introduction}

In the dense environment of circumstellar disks, grains are subject
to coalescence and fragmentation, which will, in time, alter their
properties and very likely cause significant growth
from the sub-micron size typical of dust in the diffuse interstellar
medium (ISM).
This is a necessary step for all theories of planet formation.
However, the interplay of  a multiplicity of poorly known
physical processes and their
dependence on details of the grain structure make it difficult to reach
consensus on the extent and timescale of dust processing
(see Beckwith et al.~\cite{Bea00} and references therein).
It is therefore important to derive constraints from the observations,
determining 
the properties of dust in disks surrounding
pre-main--sequence stars of well known ages.

The  spectral energy distribution (SED) of T Tauri stars at millimeter
and submillimeter wavelengths has a shallow dependence on
wavelength, consistent with optically thin emission from
millimeter-size grains (Beckwith \& Sargent \cite{BS91}).
However, this interpretation of the SEDs 
has remained controversial,
since it was soon realized that it is not possible to infer 
the dust opacity law from the
SED alone, and that  the  effect of potentially large optical depth
needs to be sorted out.
To do that, one
needs to combine the determination of the long-wavelength SED
with spatially resolved
images of the disk at one (or possibly more)  wavelengths
 (see e.g. the discussion in Testi et al.~\cite{Tea01}).
Recently, this  has been done for two pre-main--sequence stars,
TW Hya  (Calvet et al.~\cite{Cea02}) and CQ Tau (Testi et al.~\cite {Tea03}),
using the VLA array at 7~mm to resolve the disk emission.
In both cases, it was found that the dust opacity depends on wavelength
roughly as $\lambda^{-0.6}$, and this was
interpreted as evidence for
grain growth  to centimeter sizes.

Following our work on CQ Tau,
we have started a search for pre-main--sequence stars of intermediate mass with 
shallow SEDs
extending to 7~mm. The choice of this rather long wavelength was
motivated by the advantages of using the largest
wavelength range for the determination of the spectral index and by the
fact that the VLA offers the best resolution available today, provided that
the source is strong enough. In addition, the longer 
the wavelength the more severe  the constraints on dust properties
one can set.

We report in this paper the results for  a sample of six stars. The sample and the
observations obtained with the PdB and VLA interferometers  are described
in \S 2.  The results are presented in \S 3 and their implications for 
grain properties and evolution are discussed in \S 4.

\section {Observations}

\subsection {Sample Stars}
Our sample  is composed by nine  pre-main--sequence stars, 
whose properties are summarized in Table
\ref{stars} (Column 1--7). 
The first six stars listed are new observations, while the last three
are from the literature (Testi et al.~\cite{Tea01},\cite{Tea03},
Calvet et al.~\cite{Cea02}). The six new stars were
selected because they were known to have shallow spectra at sub-millimeter or millimeter wavelengths from previous observations
(Sylvester et al.~\cite{Syea96}, Mannings \& Sargent \cite{MS97},\cite{MS00}). 
We adopted distances and spectral types from Hipparcos and computed the
stellar parameters (effective temperature and luminosity) as  in Natta et al.~(\cite{Nea00}). Slight differences from the
values obtained by other authors are within the uncertainties; they
are not significant for the purposes of this paper.
For HD~34282, we have adopted the recent estimate of the 
distance and luminosity of Pi\'etu et al.~(\cite{Pietu03}).
HD~35187 is a binary with separation of 1.38$''$; following 
Dunkin \& Crawford (\cite{DC98}),
we have attributed the disk emission to the primary component HD~35187B. HD~143006 has no measured Hipparcos distance; we have adopted the distance of 82 pc estimated by Sylvester et al.~(\cite{Syea96}), as well as their spectral type.

Stellar masses and ages are estimated from the location of the stars on the HR diagram,
using Palla \& Stahler (\cite{PS93}) evolutionary tracks.
The stars in the sample are all relatively old objects
($>$few million years), as expected given their ``isolated''
nature. They cover a  range of masses from about 1 to 2.3 solar masses, and luminosities from 0.23 to 42 \Lsun.

\begin{table*}
\begin{flushleft}
\caption{ Star and Disk Properties}
\vskip 0.1cm
\begin{tabular}{lccccccccccc}
\hline\hline
Star$^a$   & ST&  \Tstar& \Lstar& D&  \Mstar& Age & $\lambda_{max}$&$\alpha_{mm}$& $\beta$ &\Mk$^b$& $\beta_{SD}$\\
       &   &   (K)&    (\Lsun)& (pc) & (\Msun)& (Myr)&  (mm)& & (\Msun) &\\
\hline
\underline{HD~34282}& A0& 9800& 36& 400 & 2.3& 3 &  3.2 &3.0&1.3 & 0.09& 1.3\\
HD~150193&A1& 9500& 21& 160 & 2.3& 6 &7.0& 4.0&1.6& 0.02& 1.0\\
\underline{HD~163296}& A1& 9500& 36& 122& 2.3& 5  & 7.0&2.6& 0.8& 0.05 & 1.0\\
HD~35187& A2& 9100& 34& 150& 2.2& 5  &  3.6 &2.6&0.7& 0.003&1.5   \\
UX Ori &A3 &8600& 42&450 & 2.2& $3$&  7.0& 2.0&0.0& 0.03&0.1\\
HD~142666& A8& 7600& 8& 116& 1.6& $>$10&7.0&2.2&0.4& 0.01 &0.7  \\
\underline{CQ Tau} &F2 &6900& 4&100 & 1.5& $>10$&  7.0& 2.4&0.5& 0.02&0.7\\
HD~143006& G6   & 5770& 0.8& 82 & 1 &$>$10 & 3.1&2.5& 0.8& 0.005&1.2  \\
\underline{TW Hya} &K8& 4000&0.23& 55& 0.6& 10&  7.0& 2.3&0.7& 0.03&0.7\\
\hline
\hline
\label{stars}
\end{tabular}
\end{flushleft}
$^a$ Disks resolved at millimeter wavelengths are underlined.
\newline $^b$ \Mk=\Mku.
\end{table*}

\begin{table}
\begin{flushleft}
\caption{VLA Observation Summary}
\vskip 0.1cm
\begin{tabular}{lcccc}
\hline\hline
Star   & $\alpha$ (J2000)&$\delta$ (J2000)& $\lambda$& F$_\nu$\\
       &   & & (mm)&  (mJy)\\
\hline
HD~34282&05:16:00& $-$09:48:35.4 &36 &$<$0.2$^a$\\
        &        &               &7& $<1^a$\\
HD~35187& 05:24:01& 24:57:37.6& 36&1.0 $\pm$0.03\\
        &         &           & 7&1.0$\pm$0.2\\
HD~142666& 15:56:40& $-$22:01:40.0 & 36& $<$0.2$^a$\\
         &         &             & 7& 1.7$\pm$0.2 \\
HD~143006& 15:58:37&$-$22:57:15.0& 36& $<$0.15$^a$ \\
         &         &            & 7& $<$0.7$^a$\\
HD~150193&16:40:18&$-$23:53:45.2& 36& 0.2$\pm$0.02\\
         &         &            & 7& 0.7$\pm$0.2\\
HD~163296&17:56:21 & $-$21:57:21.9& 61$^b$& 0.25$\pm$0.04\\
         &         &            & 36$^b$& 0.42$\pm$0.025\\
         &         &            & 13$^b$ & 0.9$\pm$0.25\\
         &         &            & 7 & 6$\pm$0.5 \\
\hline
\hline
\label{tobs_vla}
\end{tabular}
\end{flushleft}
$^a$) Upper limits are 3$\sigma$;
$^b$) Data obtained from VLA Archive, originally observed by Bouwman et al.~(\cite{JBea00})
\end{table}
\begin{table}
\begin{flushleft}
\caption{PdB Observation Summary}
\vskip 0.1cm
\begin{tabular}{lcccc}
\hline\hline
Star   & $\alpha$ (J2000)&$\delta$ (J2000)& $\lambda$& F$_\nu$\\
       &   & & (mm)&  (mJy)\\
\hline
HD~34282&05:16:00& $-$09:48:35.4 &3.2 &5.5$\pm$1\\
        &        &               &1.3& 100$\pm$23\\
HD~35187& 05:24:01& 24:57:37.6&  3.6& 2.5$\pm$0.4\\
        &         &           &  1.3& 20$\pm$2\\
HD~142666& 15:56:40& $-$22:01:40.0  & 3.3& 11$\pm$1\\
         &         &             &  3.1& 13$\pm$1\\
         &         &             &  1.2& 79$\pm$4\\
HD~143006& 15:58:37&$-$22:57:15.0&  3.1 &4.6$\pm$0.7 \\
         &         &            &  1.2 &43$\pm$3\\
\hline
\hline
\label{tobs_pdb}
\end{tabular}
\end{flushleft}
\end{table}

\subsection {VLA  observations}

New NRAO\footnote{The National Radio Astronomy Observatory is a facility
of the National Science Foundation operated under cooperative agreement
by Associated Universities, Inc.} VLA 7~mm and 3.6~cm observations of 
HD~34282, HD~35187, HD~142666, HD~143006, HD~150193, and HD~163296
(for the latter only 7~mm data was obtained) were performed
on several occasions from November 2001 to May 2003. The array was 
mainly used in the most compact D configuration, offering baselines
from the shadowing limit to $\sim$350~m; the corresponding
angular resolution at 7~mm is $\sim 2$ arcsec. HD~142666,
HD~143006, HD~150193, and HD~163296 were 
also observed in the C configuration, with baselines up to 3.4~km
and resolution $\sim 0.5$ arcsec.
In both cases the largest angular scale that can be accurately imaged
is $\sim 40^{\prime\prime}$.
Calibration and editing of the ($u,v$) data were performed using standard 
techniques within the AIPS package. Images were produced using the
AIPS IMAGR task with natural weighting of the visibilities. 
The measured fluxes or upper limits are reported in Table~\ref{tobs_vla}.
For HD~163296, there were previous data at centimeter wavelengths
in the VLA archive (see also Bouwman et al.~\cite{JBea00});
we have re-reduced them with the same procedure used for
our new observations and added the resulting fluxes to Table~\ref{tobs_vla}.
Flux calibration was performed using standard techniques at 3.6 and 0.7~cm
observing 3C286 and/or 3C48 and deriving the flux density at the time 
of observations for the phase calibrators. This procedure is expected
to be accurate within 1\% at 3.6~cm and within $\sim$15\% at 0.7~cm.

\subsection {Plateau de Bure observations}

Observations of four of our target stars were made simultaneously at $\sim 3$~mm and
1.2~mm using the Plateau de Bure interferometer 
\footnote{The Plateau de Bure interferometer at IRAM is supported by
INSU/CNRS (France), MPG (Germany) and IGN (Spain).}
on May 18 and 29, 2001 and August 18 and 19, 2001.  Visibilities were obtained
in the most compact configuration of the 5 antenna array, yielding projected
baselines which range from about 64 m down to the antenna diameter of 15 m.
The largest angular scales that can be imaged range from
$\sim$20$^{\prime\prime}$ at 3.5~mm to less than 8$^{\prime\prime}$ at 1.2~mm.
The $46''$ ($20''$ at 1.2~mm) primary beam field of the interferometer was
centered at the nominal position of the each star (see Table~\ref{tobs_pdb}).

At 1.2~mm, data were taken in double sideband mode, while at $\sim 3$~mm observations
were made in upper sideband only (see Table~\ref{tobs_pdb} for the tuning
frequencies).  


Data calibration was performed in the antenna-based manner. Flux densities
were obtained from the visibilities using standard IRAM fitting procedures. 
The calibration is expected to have an uncertainty of $\sim$15\% at 3~mm and
$\sim$20\% at 1.2~mm.
All the sources turned out to be unresolved;
the derived fluxes are given in Table~\ref{tobs_pdb}.


\section {Results}

\begin{figure*}
\begin{center}
\leavevmode
\centerline{ \psfig{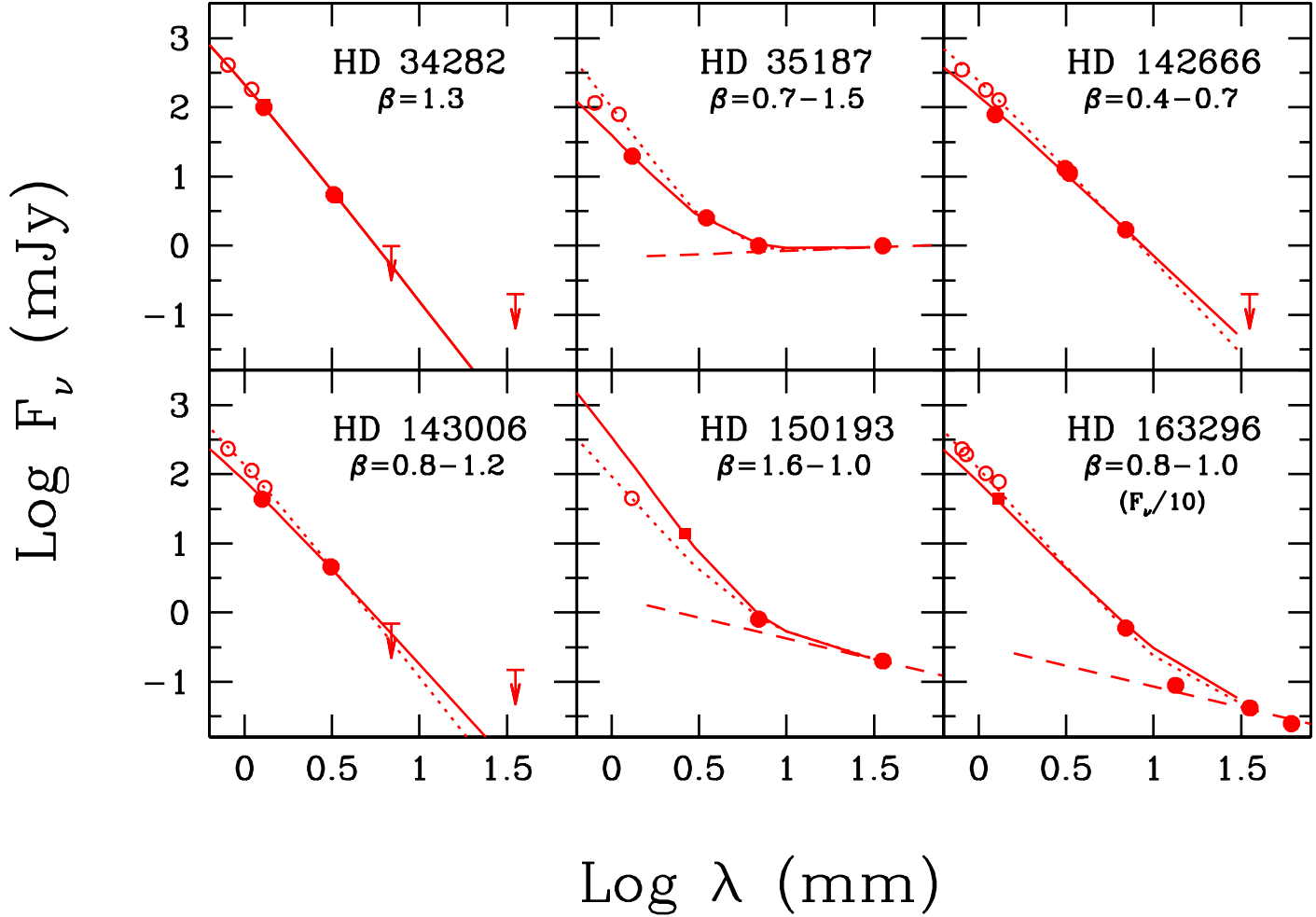} }
\end{center}
\caption{Observed fluxes for the six stars in our sample.  Filled dots and
arrows show detections and 3$\sigma$ upper limits from this paper (PdB and
VLA); filled squares are interferometric data from  Pi\'etu et
al.(\cite{Pietu03}; HD~34282, nearly
coincident with our points) and  Mannings and Sargent (\cite{MS97}; HD~150193
and HD~163296), respectively. Open dots are single dish JCM data (Sylvester et
al.~\cite{Syea96} and Mannings~\cite{M94}).  Dashed lines show the adopted fit
to the centimeter fluxes; The solid lines show the results of disk models, to
which we have added the estimated free-free contribution (see text).  Dotted
lines show disk models that fit the single-dish fluxes (rather than the
interferometric ones) at $\lambda \le 1.3$~mm.  The corresponding values of
$\beta$, the opacity power-law exponent, are shown on each panel; the first
value  has been derived fitting our  interferometric data at all wavelengths
($\beta$ in the text and in Table~2), the second using single-dish data for
$\lambda \le 1.3$~mm (\bsd\ in the text and in Table~2).
Note that for HD~163296 the models overpredict the 1.3 cm flux; this happens
because we have assumed that the dust opacity is an unbroken
power-law to $\lambda=3$ cm, which
is likely not the case.
}
\label{mm}
\end{figure*}

Fig.~\ref{mm} shows  the observed fluxes as function of wavelength for
each star.
PdB and VLA detections are shown as filled dots; the  errors are
smaller than the symbol size. Arrows are $3\sigma$ upper limits.

We show in Fig.~\ref{mm} also interferometric data from the literature, obtained with 
PdB  (Pi\'etu et al.~\cite{Pietu03}) and OVRO
(Manning \& Sargent \cite{MS97}; \cite{MS00}), as well as single-dish fluxes
measured with the JCMT by Sylvester et al.~(\cite{Syea96}) and 
Mannings (\cite{M94}).
At about 1~mm, where the two sets of data overlap,
the single-dish fluxes tend to be larger than the  interferometric ones by factors of order 1.3--1.6, somewhat larger  than the calibration
errors (which are of the order of 20\%). Note however that this is not always
the case, e.g. HD~34282 
where the single-dish and interferometric fluxes
agree within the errors.
It is likely that this 
difference is caused by
 calibration uncertainties and by the different width of the
interferometer and single-dish bolometer
bandpass , but it is also possible that the interferometers
are missing some diffuse emission. We will go back to this point in \S 3.3.

\subsection {Radio fluxes and gas emission}

The main purpose of our 3.6cm observations was to check for contamination of the millimeter
dust emission by free-free  or non-thermal emission.
Three  objects in our sample were not detected at 3.6cm:
HD~34282, HD~142666 and HD~143006; two of these
(HD~34282 and HD~143006) were not detected at 7~mm either.
The upper limits shown in Fig.\ref{mm} indicate
that at millimeter wavelengths
there should be very little contribution  from gas emission.

Three  objects were  detected. HD~35187 is a very strong radio source;
the spectral index between 7~mm and 3.5~cm is $\sim 0$. 
It is possible that we have detected a radio flare; in any case,
we have made the conservative assumption that  
only a negligible fraction of the
observed  7~mm flux is  due to dust, and evaluated the ionized
gas contribution to
shorter wavelength   as
a power-law of slope -0.1 (dashed line in  Fig.~\ref{mm}).

HD~150193 has a rather steep slope between 7~mm and 3.5~cm. We have assumed that the 3.5cm flux is
free-free emission from a wind, for which we expect a spectral slope
of $\sim 0.6$ (Felli \& Panagia \cite{FP81}). The dashed line
in Fig.~\ref{mm} shows the estimated gas emission; it accounts for
$\sim 70$\% of the observed 7~mm flux.

In the case of HD~163296 there are additional radio observations
at 1.3cm and 6cm (Bouwman et al.~\cite{JBea00}). The three
centimeter-wavelength fluxes
are consistent with wind emission with slope $\sim 0.6$  
(dashed line in Fig.~\ref{mm});
the corresponding contribution at  7~mm is negligible.

\subsection {Dust emission}

Table~\ref{stars}, Column 9 shows the spectral index
of the dust emission
\amm\ ($F_\nu \propto \nu^{\alpha_{mm}}$),
 after subtraction of the free-free component, measured
between 1.3~mm and a maximum wavelength (either 2.6 or 7.0~mm) given
in Column 8. The error on \amm\ is  about $\pm 0.2$ and is
dominated by the systematic uncertainties on the flux scale
at different wavelengths and telescopes.
In all objects but one \amm$\simless 3$. The exception 
is HD~150193, which has  no interferometric measurement at 1.3~mm;
\amm\ is  computed  between 2.6 and 7~mm, and is therefore much more
uncertain than in the other objects.

\subsection {Disk models}
Relatively flat spectral indices at
millimeter wavelengths can be the signature of optically thin emission
from grains whose opacity has a very shallow dependence on wavelength,
but also of optically thick emission from grains of any kind.
Once all other parameters are fixed, it is possible to obtain
shallower and shallower values of \amm\ down to the limit \amm=2 by
increasing the disk mass (and therefore making the dust emission
optically thick). However, at any given wavelength the corresponding
flux will increase to very large values, unless the disk is
very small and highly tilted to the line of sight (see the discussion
in Testi et al.~\cite{Tea01}).

A good illustration is the case of UX Ori, which has the flattest
spectral index so far (\amm=2.03$\pm$0.26; Testi et al.~\cite{Tea01}).
The observations could be reproduced by disks with ISM grains if
\Rd =30~AU, $\theta$=66\grad, \Md$\sim$2 \Msun.
Larger disks need to be more  tilted
($\theta=$86\grad\ for \Rd=100 AU, for example), obscuring the central star,
and even more  massive, well above the limit where disks become unstable to self-gravity.
If the disk is large (say, $>$100 AU), then
it must be optically thin at millimeter wavelengths;
the observed \amm\ can only be reproduced if the dust opacity is practically independent of wavelengths.

We have analyzed the stars in this sample following the same procedure 
already used for UX Ori and CQ Tau by Testi et al.~(\cite{Tea01}).
We  derive the dust opacity law by comparing
for each star the observed fluxes, corrected for  free-free
contamination, to the predictions of self-consistent disk models.
We use  the two-layer models of flared disks (i.e., in hydrostatic
equilibrium) heated by stellar irradiation
as developed by Dullemond et al. (\cite{DDN01}), 
following the schematization of Chiang \& Goldreich (\cite{CG97}).
These models have been used in the analysis of CQ~Tau by 
Testi et al.~(\cite{Tea03}) and we refer to that paper for
a more detailed description.

To characterize completely such a disk model,
once the stellar properties are known, one needs to specify a
number of parameters, namely the inclination $\theta$ with respect to the
observer ($\theta=0$ for face-on disks), the disk inner and outer radii
(\Ri\ and \Rd), the disk mass (\Md),
the dependence of the surface density on radius
($\Sigma \propto R^{-p}$), the properties of dust on the disk surface and in the
midplane. However, millimeter fluxes are practically independent of some
of these parameters, for example the disk
inner radius and the surface dust properties. Also, they are insensitive
to most details of the adopted radiation transfer scheme (see
Dullemond \& Natta \cite{DN03}).

For each star  we have computed a large grid of models, varying
\Rd, \Md, $p$, $\theta$ and the dust opacity in the midplane, 
described by a power-law 
$\kappa=\kappa_{\rm{1mm}}(\lambda/\rm{1mm})^{-\beta}$
with $\kappa_{\rm{1mm}}$=0.01 cm$^2$ g$^{-1}$  as a fiducial
value (which includes the gas contribution to the total mass
for a gas-to-dust mass ratio 100). 
\Ri\ is fixed at the dust sublimation radius; for the 
properties of the 
dust on the disk surface we have adopted the same model as Testi et al.~(\cite{Tea03}); neither of these
model parameters  is constrained by our data, nor do these choices affect
the interpretation of the millimeter spectra.

For two of the stars in the sample, the millimeter observations
resolve the disk emission and provide lower limits on \Rd.
HD~34282 has been resolved at 1.3~mm by Pi\'etu et al.~(\cite{Pietu03}), who measure
a deconvolved FWHM size of $1.74\pm 0.07''\times 0.89\pm 0.06''$ and derive an inclination of $56\pm 3$deg.
Note that the fluxes measured by Pi\'etu et al.~(\cite{Pietu03}) agree
with our measurements to within the errors.
Since the deconvolved FWHM size is a lower limit to the physical size
(Dutrey et al.~\cite{Dea96}; Pi\'etu et al.~\cite{Pietu03};
Testi et al.~\cite{Tea03}),
at a distance of 400 pc, the outer disk radius must be
$R_d> 350$ AU.
With this lower limit to  $R_d$ and the observed
$\theta$, our models fit the observed fluxes
only if the disk is optically thin at millimeter wavelengths. 
Thus, 
the only  quantities that
one can derive from the data are $\beta$ and the disk mass for the
fiducial value of $\kappa_{\rm{1mm}}$, that we will call
\Mk=\Mku. The results are
$\beta \sim 0.7$ and
\Mk$\sim 0.1$ \Msun, respectively. 
The  uncertainties are  typically 
 about $\pm 0.1$ for beta and a factor of 2
for \Mk\ (for $p=1-2$), as discussed in Testi et al.~(\cite{Tea03}). 
They are due to the fact that the surface density profile is not
well constrained by the data, and  it is possible to reproduce the same
intensity map with  different values of $p$ by changing \Rd\ (the larger \Rd,
the steeper the density profile one needs).
Note that for HD~34282 the value of $\beta$ would be the same if the
distance were D=160 pc, as measured by Hipparcos, but the
disk mass would be reduced by a factor  $\sim$6.

HD~163296 is partially resolved in 1.3~mm continuum
emission by Mannings \& Sargent (\cite{MS97}), 
who measure a deconvolved FWHM size of 110$\times$95 AU;
in the CO (2-1) line the disk size is 310$\times$160 AU.
Our 7~mm observations with the VLA in C configuration
indicate that the emission is  resolved with deconvolved size  about
300$\times$180 AU and
inclination 55$^\circ$ (Testi et al.~\cite{Tea04}). As soon as
\Rd$\simgreat 100$ AU,
the millimeter emission is optically thin 
with $\beta\sim 0.8$ and \Mk$\sim$0.05 \Msun.

To the best of our knowledge,
none of the other disks have been resolved at any wavelength, and we have no indication of the disk outer radius.
However, we consider it unlikely that they are all very small, massive disks,
seen almost edge-on, as one would require to reproduce the observations
if the grains are as in the ISM.
We have therefore  proceeded assuming
\Rd$>$100 AU and allowing the other disk parameters to vary, 
including the dust opacity. 
The resulting values of $\beta$ and \Mk\ are shown in Table \ref{stars},
Column 9 and 10, respectively.
Note that a small,
truncated disk may be more likely in HD~35187, which is a binary
with separation of 200 AU. Still, a disk with \Rd$\sim$ 70 AU (i.e., 1/3 the
separation) reproduces the observed fluxes only if it is optically thin.

The unresolved fluxes analyzed  here do not constrain the value
of $p$, since it is generally possible to fit the SED data with
either very flat ($p=0.5$) or very steep
($p=2$) surface density profiles. Increasing $p$ increases the contribution
to the millimeter fluxes of the inner, optically thick disk, whose
emission does not depend on $\beta$, and we may wonder how much mass
can be "hidden" in this region. However, we find that
\Mk\ is a rather good estimate of the total mass of the grains that
contribute to the millimeter-wavelength opacity.
It should be noted however, that all available high angular resolution
observations suggest that $p\le1.5$ (Dutrey et al.~\cite{Dea96}; Wilner 
et al.~\cite{Wea00}; Kitamura et al.~\cite{Kea02}; Testi et al.~\cite{Tea03}).
Typically, in disks with flat surface density profile ($p \simless 1.5$), the
inner, optically thick (at $\sim$1.2~mm) part of
the disk contains only about 10\% of the total mass.

One final potential source of uncertainties on our estimates of $\beta$ stems
from the characteristics of the interferometers themselves, which filter out
any emission more extended than the largest angular scale mentioned in 
Sect.~2.2; in particular, at 1.2~mm, this effect could be important.
To estimate the effect of this
potential problem, we have computed values of $\beta$ (\bsd)  fitting the
single-dish fluxes  at $\lambda \le 1.3$~mm and the longer wavelength
interferometric points (Column 11). This combination gives a steeper spectral
slope, and a correspondingly higher value of $\beta$. This value is likely to
be largely overestimated.  In fact, if there is a diffuse overlying emission
component, this should also be present at longer wavelengths, and by combining
in this way  single-dish and interferometer data we are biasing our results
toward steeper spectral slopes.  Additionally, if diffuse emission is indeed
present, it is unlikely to come from the disk itself, so that we should in any
case exclude it when fitting disk models to the data.  In spite of these
considerations, one shoud keep in mind the possibility that the interferometric
fluxes are somewhat underestimated, and the values of \bsd\ provide a rough
estimate of the observational uncertainties in the derivation of $\beta$.

\section {Discussion}

The results of our analysis are summarized in Table \ref{stars}. To
the six stars in our
sample, we have added UX Ori and CQ Tau (data from Testi et al.~\cite{Tea01},
\cite{Tea03} and references therein)
and TW Hya (data from Weintraub et al.~\cite{W89},
Wilner et al.~\cite{Wea00},\cite{Wea03} and references therein), and we have  reanalyzed them again in the same manner
for homogeneity. 
The objects that have been spatially resolved are underlined.
In all the objects the values of \bmm\ are well below 2, which is the
typical wavelength dependence of the opacity of
small ISM grains. This is true also if we consider the
upper limits \bsd.
For the four resolved systems, the  disk sizes are such that optically thick
emission is not a viable possibility (see \S 3.2 and Testi et al.~\cite{Tea01},
\cite{Tea03}, Calvet et al.~\cite{Cea02}). There is no doubt that in
these objects the grains have been hugely processed. The remaining
five systems have not been resolved; however, as discussed in \S 3.2,
although the values of $\beta$ we derive for these objects are more
uncertain (because the observations do not exclude an optically thick
contribution to the emission),  
it is very unlikely that all of these stars have very compact disks
with  unprocessed ISM grains.

Once we have established that  the grains originally present in the
disk have undergone a large degree of processing, can we make the
next step and derive
the properties that characterize the actual grain 
population, for example the maximum grain
size? This is not straightforward when considering the complexity added
by a wide distribution of grain sizes and composition, as we will now illustrate.

\begin{figure}
\begin{center}
\leavevmode
\end{center}
\centerline{\psfig{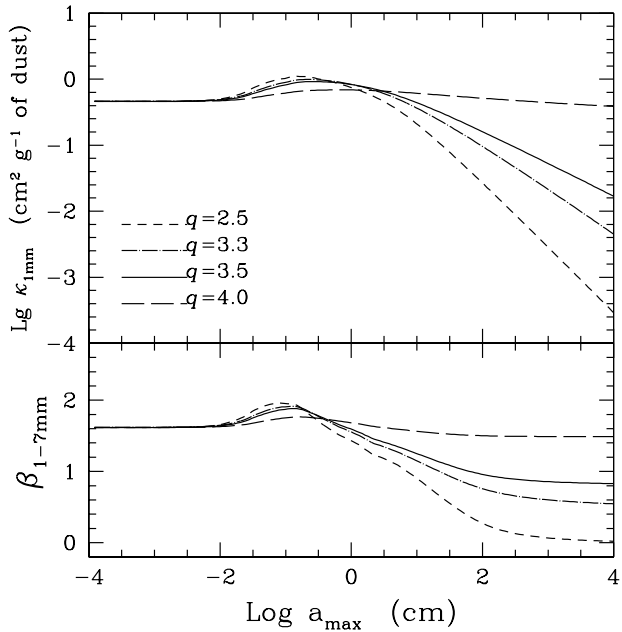} }
\caption{Top Panel: dust opacity at 1~mm as function of \amax\ for a size
distribution $n(a)\propto a^{-q}$ between \amin=0.01$\mu$m and \amax.
Different curves are for different values of $q$, as labelled.
The grains are porous conglomerates of 5\% (in volume) olivine 
([Fe$_{0.3}$Mg$_{0.7}$]$_2$SiO$_4$), 15\% organic materials,
35\% water ice and 50\% vacuum; see 
$http://www.astro.uni-jena.de/Users/dima/Opacities/RI/new\_ri.html$
for the cross sections of the individual components. Cross sections of the
porous conglomerates have been computed as in Kr\"ugel and
Siebenmorgen~\cite{KS94}.
Bottom Panel: \bmm\ between 1~mm and 7~mm as function of
\amax\ for the same grain distributions.
}
\label{beta}
\end{figure}

Let us consider a population of grains with a power-law size distribution
$n(a)\propto a^{-q}$ between a minimum and a maximum
size, \amin\ and \amax, respectively. 
In the ISM, \amin\ is few tens of \AA, \amax\ is $\sim 0.1-0.2$ \um, and $q=3.5$. 
These values will change if grains are processed in disks, and one can expect
much larger values of \amin, \amax, and a variety of values of $q$. 
Small values of $q$ are expected when coagulation processes dominate,
while large values of $q$ characterize fragmentation (see, for example,
Weidenschilling~\cite{W97}).
For a given dust model, i.e., once the chemical structure, composition, porosity etc. are specified, one can  compute
the opacity for different values of \amin, \amax\ and $q$.
We choose for our discussion porous composite grains with 50\% vacuum
and state-of-the-art cross sections, as detailed in the
caption of
Fig.~\ref{beta}. 
The top panel shows the average dust
opacity at 1~mm as function of \amax\ for various values of $q$; the bottom panel the corresponding values of \bmm. 
The results do not depend on \amin\ as 
long as it is $\ll \lambda$. 

The value of \bmm\ that characterizes the average
opacity of  a grain distribution decreases as \amax\ increases,
as expected since grains with size $\gg \lambda$ have wavelength-independent
opacity. However, only for $q <3$, \bmm\ goes to zero
for large \amax;
for $q >3$, the small grains always contribute to the
opacity, so that, as \amax\ increases,
 $\beta$ reaches an asymptotic values that depends on $q$ and
on the  $\beta$ of the small grains. For $q$=4, the asymptotic value is
practically that of the small grains.

Three objects in our sample (CQ Tau, TW Hya, HD 142666)
have $\beta \sim 0.5-0.7$. 
This is consistent with $q\sim 2.5$ and \amax\  $\sim 80$ cm;
a distribution with $q=3.3$ has an asymptotic value \bmm=0.5,
and it is consistent with the observations as soon as \amax$>300$ cm.
In both cases the 1~mm opacity is much lower than our fiducial value, and the
estimates of the disk mass needs to be raised by a factor $>40$ for $q=3.3$ and
about 15 for $q=2.5$. The corresponding disk masses
 are high ($>$0.3, 0.9, 0.3 \Msun, respectively); however, only in the case of TW Hya
they violate 
the requirement that the disks must be gravitationally stable, i.e.,
\Mratio $\simless$30\%. 
Objects like HD~163296 have \bmm$\sim$0.8--1; they are consistent with $q=2.5$,
\amax$\sim$ 10~cm, that would results in a disk mass 0.3 \Msun,
or with any steeper size distribution with larger \amax.
A distribution with $q=3.6$ has an asymptotic value \bmm\ $\sim 1$ for
\amax\ $\simgreat 150$~cm, the corresponding disk mass would be $>$0.5 \Msun.
Very steep size distributions, such as $q=4$,  never fit the
observations.

The most extreme case in our sample, UX Ori (for which, however, there is no
spatially-resolved disk image) has $\beta\sim 0$. 
If the disk is optically thin, this would require flat
grain size distributions and very large \amax; for example,
$\beta=0.2$ corresponds to $q=2.5$, \amax$\sim 10^3$ cm. For such grain size distribution, however, 
the disk mass becomes  $\sim$ 4 \Msun, i.e., $\gg$\Mstar

%
These results depend 
on the dust model one adopts. 
For example, if the fraction of vacuum in the same grain model 
of Fig.~\ref{beta} is reduced from 50\% to 10\%, the value of \amax\ which 
reproduces the observations is reduced by
a factor $\sim 3-8$ and \Md\ by a factor $2-3$ ($q\le 3.3$). 
Grains of the same
size and composition, but different structure and topology
(homogeneous, composite aggregates, porous  homogeneous or composite
spherical particles, onion-shells particles, etc.) 
can have  different values of \bmm\ and $\kappa_{1mm}$ (see,
for example, 
Miyake and Nakagawa~\cite{MN93},
Kr\"ugel and Siebenmorgen~\cite{KS94},
 Semenov et al.~\cite{SHI03} and the Jena web page),
leading to rather different estimates of \amax\ and of the disk mass.
Just to mention two examples of how all this affects
the interpretation of the data,  Testi et al.~(\cite{Tea01}), 
fitted the same UX Ori spectrum discussed above
with ice-coated silicates of about 10 cm
radius and a disk mass in the range 0.3--1 \Msun, depending on the
grain density. 
Calvet et al.~(\cite{Cea02}) obtained a good fit to the millimeter
properties of TW Hya for $q=3.5$, \amax\ $\sim 1$~cm and \Md\ $\sim$0.1 \Msun,
using
the dust model of Pollack et al.~(\cite{Pea94}) of compact segregate
spheres where the relative 
fraction of water ice to organic materials was somewhat decreased (P.
D'Alessio, private communication).

All this shows that it is not possible to disentangle from the value
of $\beta$, which is just an average quantity, all the details that
specify a given  dust model.
However, there are some basic grain properties that are constrained,
since
no realistic grain model results in \bmm$< 1$ for \amax$< 1$~mm.
In fact, for most known grain mixtures, 
our observations are consistent with power-law grain size distributions
where \amax\ is few ten to few hundred centimeters. 

\begin{figure}
\begin{center}
\leavevmode
\centerline{\psfig{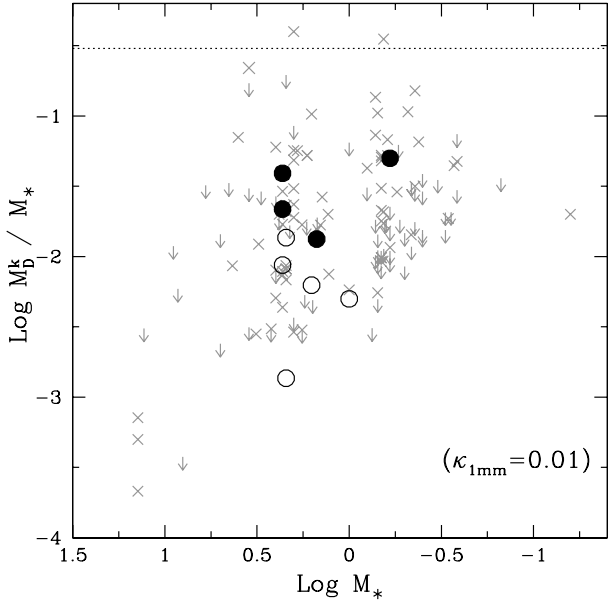} }
\end{center}
\caption{Ratio of the disk to the stellar mass as function of \Mstar for a
large sample of pre-main--sequence stars 
(adapted from Natta et al.~\cite{Nea00}). 
Crosses are detections, upper limits are indicated with arrows 
The objects studied in this paper are marked as circles;
the four resolved ones are indicated with filled circles. All the disk
masses have been computed for an opacity $\kappa_{1mm}=0.01$ cm$^2$ g$^{-1}$.
The dotted line corresponds to \Mku/\Mstar =0.3.
}
\label{diskmasses}
\end{figure}

To account for the observed fluxes, the amount of mass in these
grains has to be very large, implying that, at least when gas was
present in the standard ISM ratio, the disk was close or above the
limit for gravitational instability.
Note that we have converted the solid mass into gas+dust mass
assuming a  gas-to-dust ratio of 100. This conversion factor
may not be appropriate for the present-time disk,
if a significant fraction of the  gas has  evaporated from the
system. Still, it provides a  correct estimate of the {\it original}
disk mass, when the disk composition reflected that of the parent cloud.
It is interesting that the value  of \Mk\ of our sample
stars is similar to what is found for all the other pre-main--sequence stars
studied so far (see Fig.~\ref{diskmasses}). One wonders in  how many cases
the disk masses are underestimated by using $\kappa_{1mm}=0.01$
cm$^2$ g$^{-1}$.

Our data do not rule out the possibility that a non-negligible fraction
of the original solid mass
is in even larger bodies (km-size), from which planets can be formed.
However, it seems unlikely that
these could contain {\it most} of the solid mass, and that
the millimeter-emitting grains  were  
just the tail
of the size distribution.
If, for example, 99\% of
the solid mass is in kilometer-size bodies, whose contribution to the millimeter
opacity is negligible, their mass should be  added to the values we 
have derived, making the disk mass 100 times larger.  
%
%

An implicit assumption we have made so far is that the grain
size distribution is continuous between \amin\ and \amax.
Models of planetesimal formation (Weidenschilling~\cite{W97})
show the developement of a gap in the size distribution between $a_1 \sim$
few centimeters
and $a_2 \sim$ few meters, which tends to fill-in with time. In this case, one
should compare \amax\ with $a_1$, and  our results are consistent with these model predictions.
Still, as we have seen, it is unlikely that most of the
dust mass is on sizes $\gg a_2$, i.e., in bona-fide
planetesimals , since the solid mass with sizes $< a_1$ is already very large
(see Testi et al.~\cite{Tea03}
for a more detailed discussion of this point).
This is an interesting point, especially because all the stars
in the sample are relatively old, $\simgreat$ few million years, and in some case $\simgreat$ 10 million years.
Most models of grain evolution have a very short timescale for the formation
of planetesimals (e.g., Weidenschilling 2000 and references therein).
One possibility is that the growth of kilometer-size
bodies is in fact very inefficient, either because it is much slower than predicted or because it involves only a small fraction of the dust.
Another intriguing possibility is that the formation of planetesimals
takes place in a very early stage of the star formation, in very
massive disks
which are gravitationally unstable.
However, a discussion of these points is well behond the scope of this paper.

\section {Conclusions}

This paper presents the results of a study of six pre-main--sequence, isolated,
intermediate mass stars, performed with the PdB (at 1.3 and 2.6~mm) and
VLA (at 7~mm and 3.6~cm) interferometers. All the stars were selected to have
indications of flat spectral indexes from previous millimeter and submillimeter
data.
Our measurements confirm this indication, and  extend our
knowledge of the dust emission continuum down to 7~mm. In all the objects
\amm\ $\simless$ 3 for wavelengths between
$\sim$~1~mm  and 2.6~mm, and in 3 cases between $\sim$~1~mm and 7~mm. 

We have added to these six stars two others (UX Ori and CQ Tau), previously
studied in the same way by our group (Testi et al.~\cite{Tea01},\cite{Tea03})
and TW Hya, for which similar data exist in the literature (Calvet et al.
\cite{Cea02} and references therein).

The observations have been compared to the predictions of self-consistent
irradiated disk models to derive the properties of the millimeter-emitting
dust. As discussed by various authors,  this cannot be done
unambiguously if only the SED is known. To solve the ambiguity introduced
by optical depth effects, one needs to spatially resolve    the
emission at some millimeter wavelength, or, at least, to have an
estimate of the FWHM size. This additional information is available
for  two stars in our sample
(HD34282 and HD 163296),
and for  CQ Tau and TW Hya. For these objects, there is no doubt
that 
grains in disks are heavily processed, 
causing a net growth, to maximum sizes that
are certainly larger than few millimeters 
and, for the majority of grain models available in the literature,
 of at least few centimeters. 
The mass of the grains that have been thus processed is very large,
in some cases comparable to the mass of the central star.
Many grain size distributions are consistent with the data, but not very 
steep ones, ($q\simgreat 4$), unless micron-size grains have 
\bmm\ $\simless$ 0.5--1.

The uncertainty on \amax\ and $q$ reflects
our ignorance 
on the details of the grain chemistry, structure and topology.
However,  in spite of these  uncertainties,
we are developing a view of the outer disks of pre-main-sequence stars
as made of a huge mass of sand and pebbles, rather than of micron-size grains.
Grain growth models can and should be tested against these observations.

\begin{acknowledgements}
We thank
Endrik Kr\"ugel for having provided to us his codes
to compute the dust opacity and for very useful discussions
on dust properties.  Nuria Calvet and Lee
Hartmann for interesting comments and discussions.
This work was partly supported by  ASI grant  ARS 1/R/27/00 and ARS-1/R/073/01
to the Osservatorio di Arcetri.
\end{acknowledgements}

\end{document}